 \documentclass[final,5p,times,twocolumn,authoryear]{elsarticle}

\usepackage{amssymb}
\usepackage{lipsum}

\journal{High Energy Astrophysics}

\begin{document}

\begin{frontmatter}



\title{Different behaviors of wavelet results for type-B and type-C QPOs of MAXI J1535-571 based on NICER data}


\author[first]{X. Chen}
\author[first,second]{W. Wang}
\affiliation[first]{organization={Department of Astronomy, School of Physics and Technology, Wuhan University},
            addressline={}, 
            city={Wuhan},
            postcode={430072}, 
            state={},
            country={China}}
\affiliation[second]{E-mail: wangwei2017@whu.edu.cn}

\begin{abstract}
Wavelet analysis, in addition to power density spectra, is another method to study the quasi-periodic signals in the light curves, but has been rarely used in black hole X-ray transients. We performed wavelet analysis of X-ray timing features and quasi-periodic oscillations (QPOs) based on NICER observations of the black hole candidate MAXI J1535-571 in this paper. Separating the light curves by the confidence level of wavelet results, we find significant differences exist in the PDS, hardness ratio and mean count between light curve segments above and below the confidence level. The S-factor, which is defined as the ratio of the effective oscillation time and the total time, demonstrates distinct values between type-C and type-B QPOs. Based on our results, the S-factor for type-B QPO is very close or equal to 0, no matter the confidence level is set as 95\% or 68\%, while the S-factor of type-C QPO is significantly higher, especially in the 68\% confidence level case. We discuss the implications of the wavelet results on resolving type-B and type-C QPOs in black hole X-ray binaries.
\end{abstract}



\begin{keyword}
Black holes physics \sep X-rays: binaries \sep Stars: MAXI J1535-571



\end{keyword}

\end{frontmatter}




\section{Introduction}
\label{introduction}

Stellar-mass black holes (BHs) in X-ray binaries normally spend most of their time in a quiescent state with a low luminosity, and can only be detected during their short lasted outbursts, which usually have their flux increased by several orders of magnitude \citep{Remillard2006ARA&A..44...49R} in a few months, hence only around 70 stellar mass BHs are reported so far \citep{CorralS2016A&A...587A..61C}. A typical outburst goes through several states, showing different properties in light curves and spectra \citep{Homan2005Ap&SS.300..107H}, including the low-hard state (LHS), hard-intermediate state (HIMS), soft-intermediate state (SIMS) and high-soft state (HSS) \citep{Belloni2011BASI...39..409B}, and normally forms a "q"-shape in the hardness intensity diagram \citep{Fender2004MNRAS.355.1105F}.

Quasi-periodic Oscillations \citep[QPOs;][]{vanderKlis1989ASIC..262...27V,Nowak2000MNRAS.318..361N,Ingram2019NewAR..8501524I}) sometimes appear in the X-ray light curves during the outburst and are generally studied with the power density spectrum (PDS) in the Fourier domain. These QPOs are classified as low-frequency QPOs (LFQPOs) with central frequencies less than 30 Hz, and high-frequency QPOs (HFQPOs) with central frequencies normally greater than 60 Hz, but are less common compared to LFQPOs \citep{Belloni2014SSRv..183...43B}. The more frequently studied LFQPOs are divided into three classes depend on their properties such as the rms, the quality factor, the centroid frequency and so on, called type-A, -B and -C \citep{Wijnands1999ApJ...526L..33W,Casella2004A&A...426..587C,Casella2005ApJ...629..403C,Remillard2002ApJ...564..962R}. Type-C QPOs can appear in all states, but are mostly in the LHS and HIMS \citep{Motta2016AN....337..398M}, with high rms (up to 20\%) and wide range of centroid frequency (0.01-30 Hz). Type-B QPOs has lower rms (less than 5\%) with centroid frequency around 6 Hz. Their PDS contains weak red noise. Basically, they only appear in the SIMS, and are the characteristic that distinguishes the SIMS from HIMS \citep{Belloni2016ASSL..440...61B}. Type-A QPOs are even weaker and quite rare, and are commonly noticed in the HSS.

PDS has been widely used on QPO study for decades, and provides plenty of instructive results. However, the physical origin of QPOs is still under debate, such as the instabilities in the disk \citep{Tagger1999A&A...349.1003T}, oscillations in the corona \citep{Cabanac2010MNRAS.404..738C}, the Lense-Thirring precession of the hot inner flow \citep{Ingram2009MNRAS.397L.101I,Veledina2013ApJ...778..165V,Ingram2016MNRAS.461.1967I,Ingram2017MNRAS.464.2979I}, or the coupling of the accretion disc and the Comptonizing corona \citep{Karpouzas2020MNRAS.492.1399K, Bellavita2022MNRAS.515.2099B, Mendez2022NatAs...6..577M,Garcia2021MNRAS.501.3173G,Zhang2022MNRAS.512.2686Z}. In addition, different types of QPOs may originate with different mechanism. For example, type-B QPO are generally believed to be jet related \citep[e.g.][]{Stevens2016MNRAS.460.2796S,deRuiter2019MNRAS.485.3834D}, and have different physical origins with type-C, while type-A and type-C QPOs may share the same generation process \citep{Motta2011MNRAS.418.2292M,Motta2012MNRAS.427..595M}. Hence the QPO classification is of great importance in understanding their generation.

The PDS based QPO classification identifies QPO types based on the evolution of the rms, the quality factor, the centroid frequency, the existence of the red noise component etc. However, PDS normally are spiky and noisy, especially in the lower frequency part, which makes it hard to recognise the red noise and/or even the QPO peak. To solve this problem, the averaged PDS, i.e. averaging the PDS of each segmented light curve, is performed, which requires that the total time duration of the light curve cannot be short, otherwise there will not be enough segments to be averaged. Moreover, it assumes the PDS of each averaged segment contains the same information, which means the QPO and the band-limited noise properties cannot evolve with time. Another flaw for both PDS and averaged PDS is that, the chosen segment of the light curve cannot be very short (such as 1 s or less), otherwise the result will not be trustworthy. With all the possible flaws, sometimes it is not easy to quantitatively distinguish the QPO types \citep[e.g. the quality factor of type C QPOs, see][]{Bogensberger2020A&A...641A.101B}. 

Wavelet analysis is a common tool to study the variation of power in a given time series in many research fields such as geophysics, and has been used to study the QPO of XTE J1550$-$564 \citep{Lachowicz2010A&A...515A..65L} and IGR J17091$-$3624 \citep{Katoch2021MNRAS.501.6123K}. While Fourier transform is extremely useful and accurate in searching for periodic signals, wavelet transform can handle irregular and non-continuous signals very well, such as the fast variations of QPOs. However, many studies using wavelet or Fourier transforms have suffered from quantitatively distinguishing valid signals. Hence \cite{Torrence1998BAMS...79...61T} provided a solution for statistical significance testing that can be used in wavelet analysis. Using this method, transient QPO behaviors can be well studied \citep{Chen2022MNRAS.517..182C}. In addition, \cite{Chen2022MNRAS.513.4875C} studied the wavelet results of QPO in the source MAXI J1535-571 during its 2017 outburst with Insight-HXMT observations. Separating time series by confidence interval calculation, they found that the QPOs appear and disappear in seconds for the whole nine observations with QPO detected, which has not been observed before. This difference are also notable in PDS shape, rms, hardness ratio and spectra. Flip-flop perhaps is the most related phenomenon, even though it normally takes several tens of seconds to transit states. However, flip-flop is poorly understood, because only several sources have showed this kind of behavior so far, and they exhibit quite different properties \citep{Miyamoto1991ApJ...383..784M,Nespoli2003A&A...412..235N,Takizawa1997ApJ...489..272T,Homan2001ApJS..132..377H,Sriram2016ApJ...823...67S,Casella2004A&A...426..587C,Sriram2013ApJ...775...28S,Homan2005ApJ...623..383H,Sriram2012A&A...541A...6S,Kalamkar2011ApJ...731L...2K,Bogensberger2020A&A...641A.101B,Yang2022ApJ...932....7Y}.

MAXI J1535-571 was discovered by MAXI/GSC \citep{Negoro2017ATel10699....1N} and Swift/BAT \citep{Kennea2017ATel10700....1K} during its 2017 outburst. The following multi-band observations suggest that MAXI J1535-571 is a low-mass X-ray binary \citep{Negoro2017ATel10708....1N,Scaringi2017ATel10702....1S,Dincer2017ATel10716....1D,Russell2017ATel10711....1R} located around 4 kpc away \citep{Chauhan2019MNRAS.488L.129C}. It contains a near-maximally spinning black hole, and is perhaps viewed at a high inclination \citep{Miller2018ApJ...860L..28M,Xu2018ApJ...852L..34X,Dong2022MNRAS.514.1422D}. Different types of QPOs have been studied in this source \citep{Gendreau2017ATel10768....1G,Mereminskiy2018AstL...44..378M,Stiele2018ApJ...868...71S,Huang2018ApJ...866..122H,Bhargava2019MNRAS.488..720B}, and found a possible link between the QPO type change from type-C to type-B and the discrete jet launching \citep{Russell2019ApJ...883..198R}. The main outburst ended in May 2018, followed by several re-flares which also contains a state transition \citep{Parikh2019ApJ...878L..28P,Cuneo2020MNRAS.496.1001C}.

In this paper, we analyze the Neutron Star Interior Composition Explorer \citep[NICER;][]{Gendreau2016SPIE.9905E..1HG} observations of MAXI J1535-571 using wavelet analysis. We have improved the calculation method of QPO confidence intervals in wavelet in order to remove the influence of broad band noise (BBN). We describe this improvement in Section 2, along with the observations and data reductions. Wavelet results are shown in Section 3. In Section 4, we discuss the possible indication of the wavelet results. Finally a conclusion is made in Section 5.

\section{Observations and Data reduction}
We used all the NICER observations of MAXI J1535-571 in 2017 in this paper, and the observation IDs are listed in Table~\ref{tab:my_label}. The standard NICER processing pipeline nicerl2 and nicerl3-lc are used to process the data, while the CALDB version is xti20221001. The detectors of \#14 and \#34 are removed as recommended.

\begin{table}
    \centering
    \begin{tabular}{c|c|c}
    Observation ID	&	MJD start & Exposure	\\
    	&		&	(s)	\\
\hline
1050360103 	&	58005.30 	&	502 	\\
1050360104 	&	58008.46 	&	5405 	\\
1050360105 	&	58008.99 	&	10710 	\\
1050360106 	&	58010.00 	&	6570 	\\
1050360107 	&	58011.87 	&	1787 	\\
1050360108 	&	58012.19 	&	3617 	\\
1050360109 	&	58013.22 	&	4671 	\\
1050360110 	&	58014.05 	&	3415 	\\
1050360111 	&	58015.28 	&	2159 	\\
1050360112 	&	58016.24 	&	2888 	\\
1050360113 	&	58017.01 	&	6342 	\\
1050360114 	&	58018.37 	&	2337 	\\
1050360115 	&	58019.01 	&	9808 	\\
1050360116 	&	58020.05 	&	6718 	\\
1050360117 	&	58021.02 	&	5441 	\\
1050360118 	&	58022.24 	&	4321 	\\
1050360119 	&	58023.25 	&	4098 	\\
1050360120 	&	58020.57 	&	4142 	\\
1130360101 	&	58024.74 	&	2509 	\\
1130360102 	&	58025.76 	&	1694 	\\
1130360103 	&	58026.73 	&	3667 	\\
1130360104 	&	58027.75 	&	1839 	\\
1130360105 	&	58028.72 	&	5835 	\\
1130360106 	&	58029.75 	&	4131 	\\
1130360107 	&	58030.72 	&	4440 	\\
1130360108 	&	58031.36 	&	3838 	\\
1130360109 	&	58032.38 	&	3477 	\\
1130360110 	&	58033.03 	&	7686 	\\
1130360111 	&	58034.00 	&	7283 	\\
1130360112 	&	58035.16 	&	4931 	\\
1130360113 	&	58036.13 	&	5153 	\\
1130360114 	&	58037.03 	&	6408 	\\
1130360115 	&	58038.00 	&	1385 	\\
1130360116 	&	58065.71 	&	141 	\\

    \end{tabular}
    \caption{NICER observations of MAXI J1535$-$571 used in this paper.}
    \label{tab:my_label}
\end{table}

Wavelet analysis \citep{Torrence1998BAMS...79...61T} is used for the 1-10 keV band light curve analysis, while the time resolution is 0.0078125 s. Each NICER observation is typically divided into a dozen to several tens of Good Time Intervals (GTIs) after the data processing, with relatively long gaps between these intervals. Wavelet analysis is not capable of handling such gaps, thus it is necessary to apply wavelet analysis separately to each interval. We set Morlet wavelet as the 'mother' wavelet with $m = 6$, and the 95 percent significance level is calculated by the univariate lag-1 autoregressive \citep[see][for more technique details]{Chen2022MNRAS.513.4875C}. 

\begin{figure}
    \centering
    \includegraphics[width=0.5\textwidth]{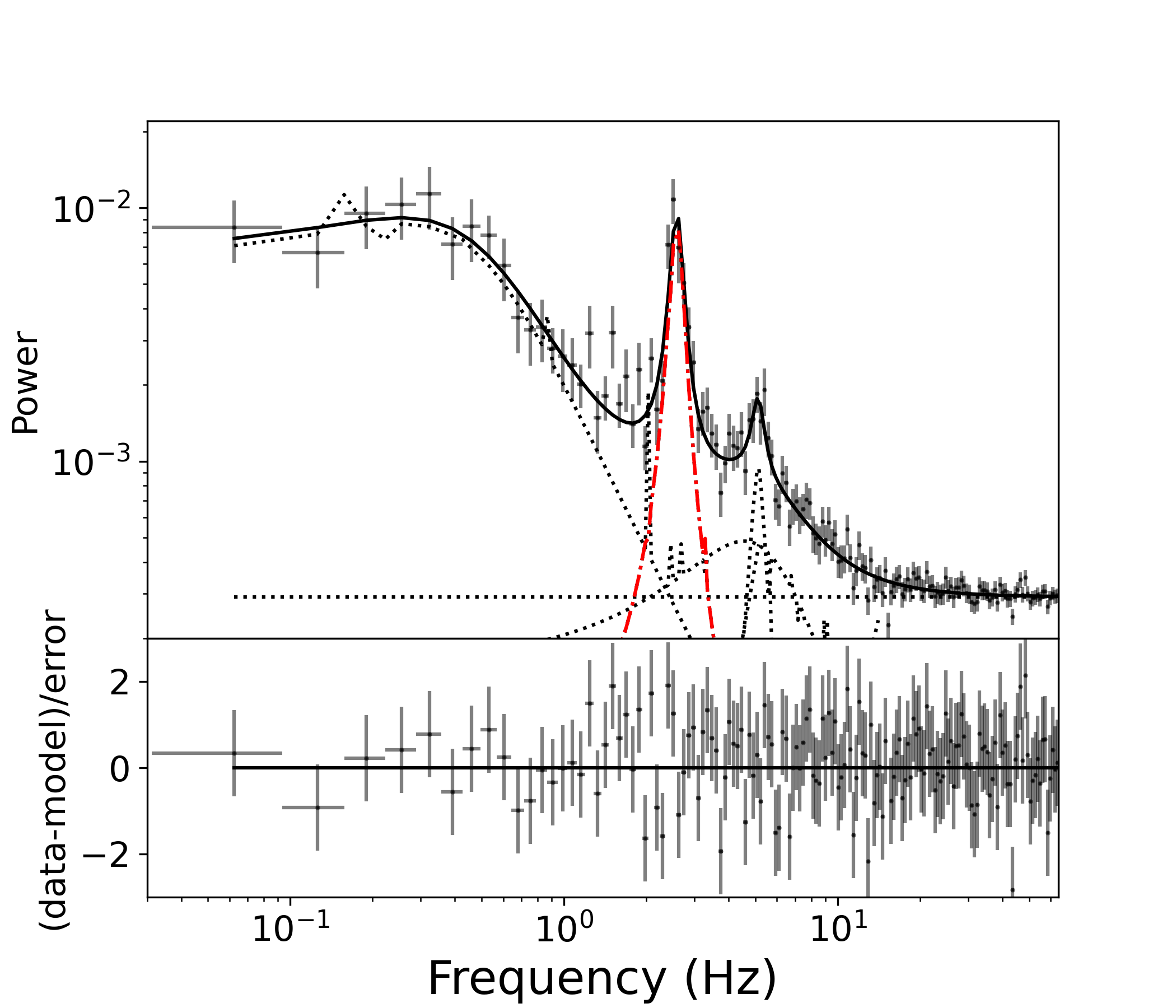}
    \caption{The 16 s averaged PDS fitting result of the first GTI in obsID 1050360104. The dashed lines show the components, while the red one marks the QPO component.}
    \label{fig:PDS}
\end{figure}

However, as can be seen from \cite{Chen2022MNRAS.513.4875C,Chen2022MNRAS.517..182C}, the BBN sometimes can also be identified as valid signals. This will obviously have an impact on the QPO confidence level calculation of the wavelet results, especially for lower energy band below 10 keV, lower QPO frequency or very weak QPO signals such as the type-B QPOs in MAXI J1535-571 as reported by \cite{Stevens2018ApJ...865L..15S} and \cite{Zhang2023MNRAS.520.5144Z}. In order to remove the influence of BBN to more accurately distinguish the time that includes QPO from the time that does not include QPO, we perform a QPO component extraction method on the wavelet results. We take the first GTI (around 220 s) in obsID 1050360104 as an example to explain our QPO component extraction method. We first calculate the averaged power density spectrum (PDS) of the GTI with 16 second time segments, with rms-squared normalization \citep{Belloni1990A&A...230..103B} been applied. We then fit this averaged PDS with multi-Lorentzian function and a constant white noise using XSPEC v12.13.0c, see Figure~\ref{fig:PDS}. After picking out the QPO component (if any, which is the red line in Figure~\ref{fig:PDS}), we calculate the proportion curve of QPO component, and apply it on the local wavelet power of the GTI, which leaves only the QPO component in the wavelet power. Meanwhile, we leave the 95\% confidence level results unchanged. We then repeat the above steps for all the other GTIs of 1050360104 and other observations.

\begin{figure}
    \begin{minipage}{0.5\textwidth}
		\includegraphics[width=1\textwidth]{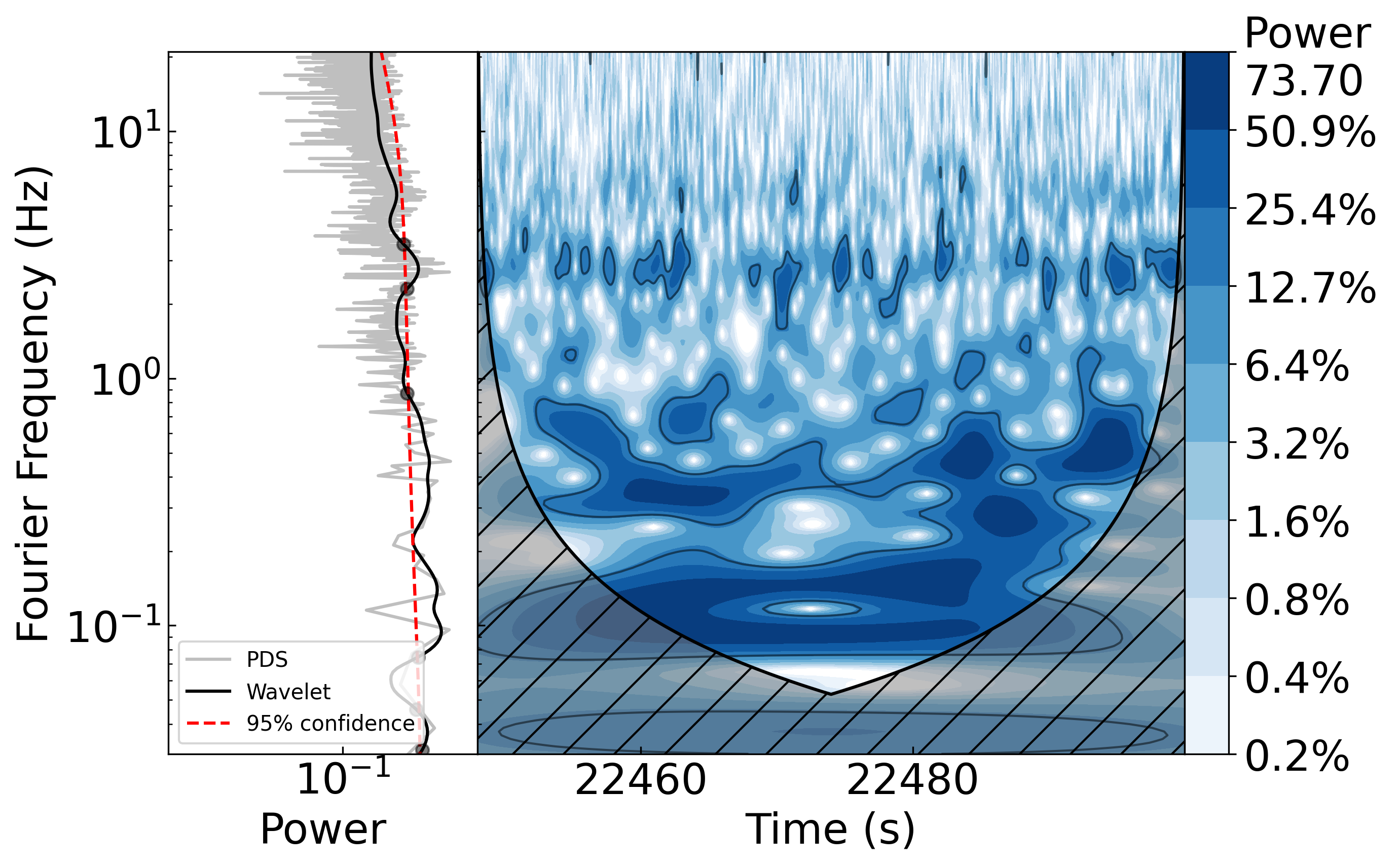}
	\end{minipage}
	\begin{minipage}{0.5\textwidth}
		\includegraphics[width=1\textwidth]{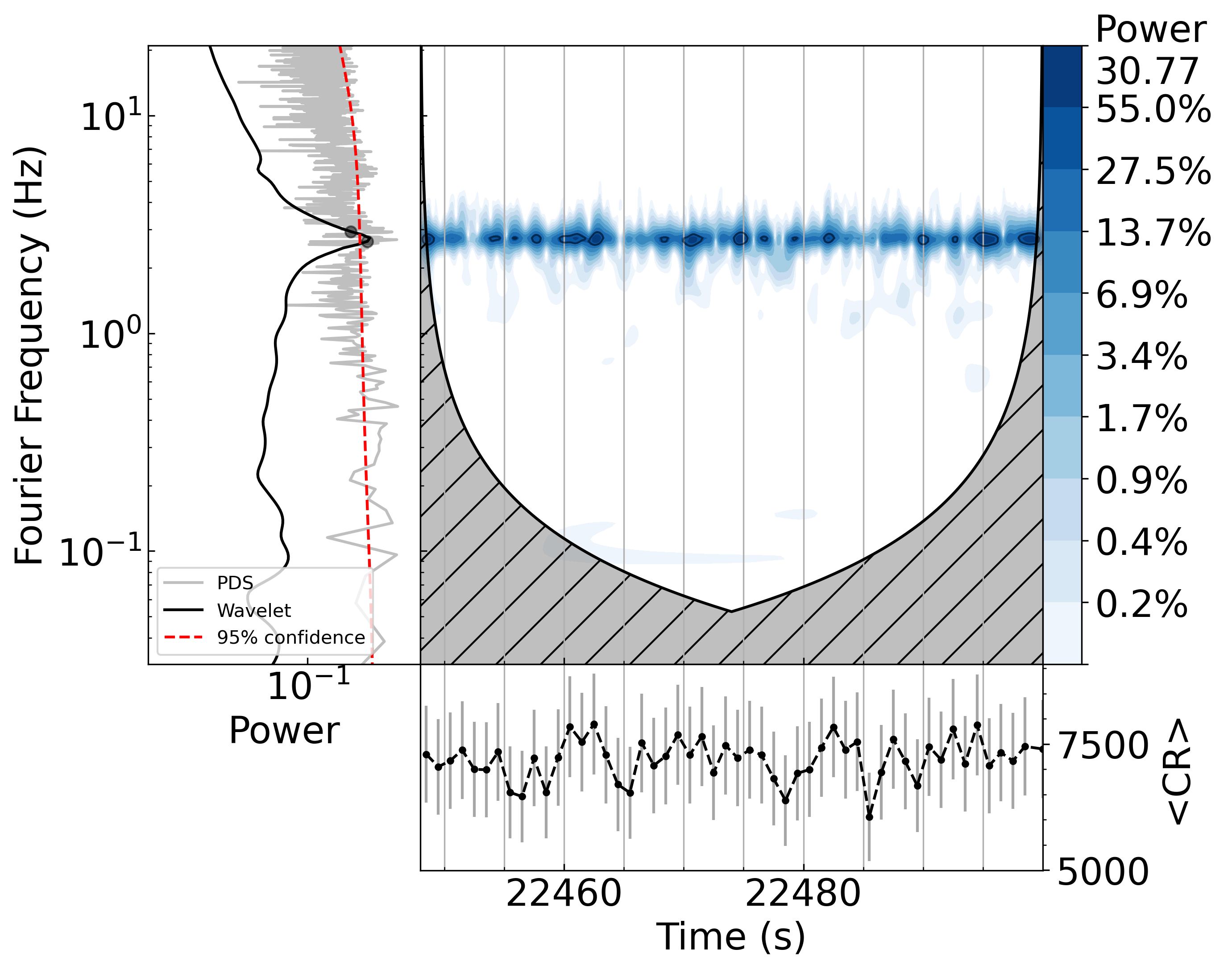}
	\end{minipage}
	\caption{Wavelet results before (top panel) and after (bottom panel) QPO component extracted. The light curve is a very short GTI from observation 1050360104 in 1-10 keV. In each sub panel, PDS (grey line), global wavelet spectrum (black line), 95 percent confidence level (red line), and local wavelet spectrum (right side) are shown. In the local wavelet spectra, the black line circled area contains significant power with confidence level greater than 95 percent, and the grey hashed area represent the cone of influence. In the color bar, we show the maximum power of the local wavelet spectrum on the top, and the underneath values are the percentage of the maximum. In the bottom panel, we add the mean count rate evolution in 1-10 keV band averaged every second.}
	\label{fig:methodResult}
\end{figure}

\section{Results}

In Figure~\ref{fig:methodResult}, we show the wavelet results before and after the QPO component extracted. It is clearly shown in the plots that after the QPO extraction process, the area exceeding the confidence level is significantly reduced, especially below 1 Hz. We present a very short interval here of around 50 s in 1050360104 to clearly exhibit the relationship between wavelet power levels and mean count rates. The circled area in the local wavelet plot, which means the power inside exceeds the 95\% confidence level, matches the "peaks" in the mean count rate very well for most of the cases, indicating a possible link between the significant power and the count rate. However, if we take the large error bars in the bottom panel into consideration, the conclusion may not be certain, but it may also explain why count rate peaks sometimes do not match the significant power. 

\begin{figure}
		\includegraphics[width=0.5\textwidth]{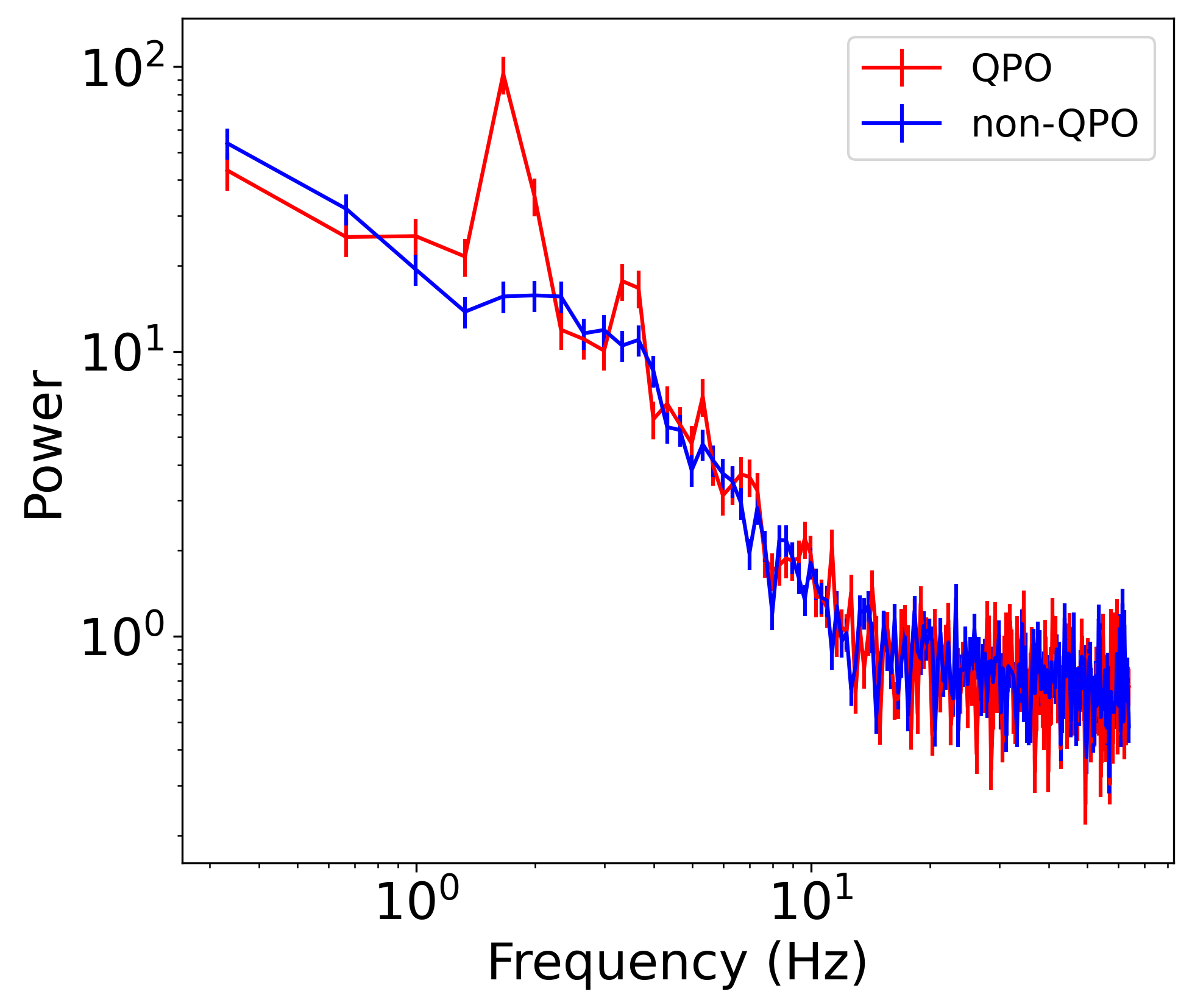}
    \caption{3 seconds averaged PDS for the first GTI in obsID 1050360106. The PDS of QPO time and non-QPO time is separated by wavelet analysis and shown with red line and blue line, respectively. Leahy normalization is performed without white noise subtracted. The wavelet is performed with 95\% confidence level.}
    \label{fig:avePDS}
\end{figure}

We compare time with significant power (hereafter QPO time) and time with insignificant power (hereafter non-QPO time) in each GTI. We notice the QPO time always have higher mean counts ($\sim$ 3 in 1-10 keV) and higher hardness ratio ($\sim 0.002$ in 7-10 keV / 1-2 keV), and the distinction can be also noticed in the PDS. For most of the observations, obvious peaks can be noticed in the QPO PDS around the QPO frequency and/or its harmonics, while the peaks almost cannot be seen in the non-QPO PDS. In Figure~\ref{fig:avePDS}, we show such an example.

\section{Discussion}
The wavelet results reveal that, the QPO signals may not appear continuously as previously thought, but appears and disappears on the order of seconds. The hardness ratio, mean count rate and PDS all show different behaviors between QPO and non-QPO time, with higher mean count rate and higher hardness ratio for the QPO time. A similar phenomenon is called the flip-flop, which also shows distinct differences in PDS and flux. However, the transition times of flip-flops are normally range from a few tens of seconds to more than a thousand seconds, which is much longer than observed in the wavelet results. It is possible that PDS cannot achieve second-level accuracy, but further research is required. In addition, type-B QPOs always appear in the flip-flops, either in the bright state or the dim state, except for Swift J1658.2-4242 \citep{Bogensberger2020A&A...641A.101B}. However, the type-B QPO is not noticed during the HIMS fast transition of this study (see Figure~\ref{fig:avePDS}), either because the time segment of several seconds is not long enough for generating a reliable PDS, or the fast transition observed in MAXI J1535-571 exhibit the same behavior with Swift J1658.2-4242.

\begin{figure}
    \begin{minipage}{0.5\textwidth}
		\includegraphics[width=1\textwidth]{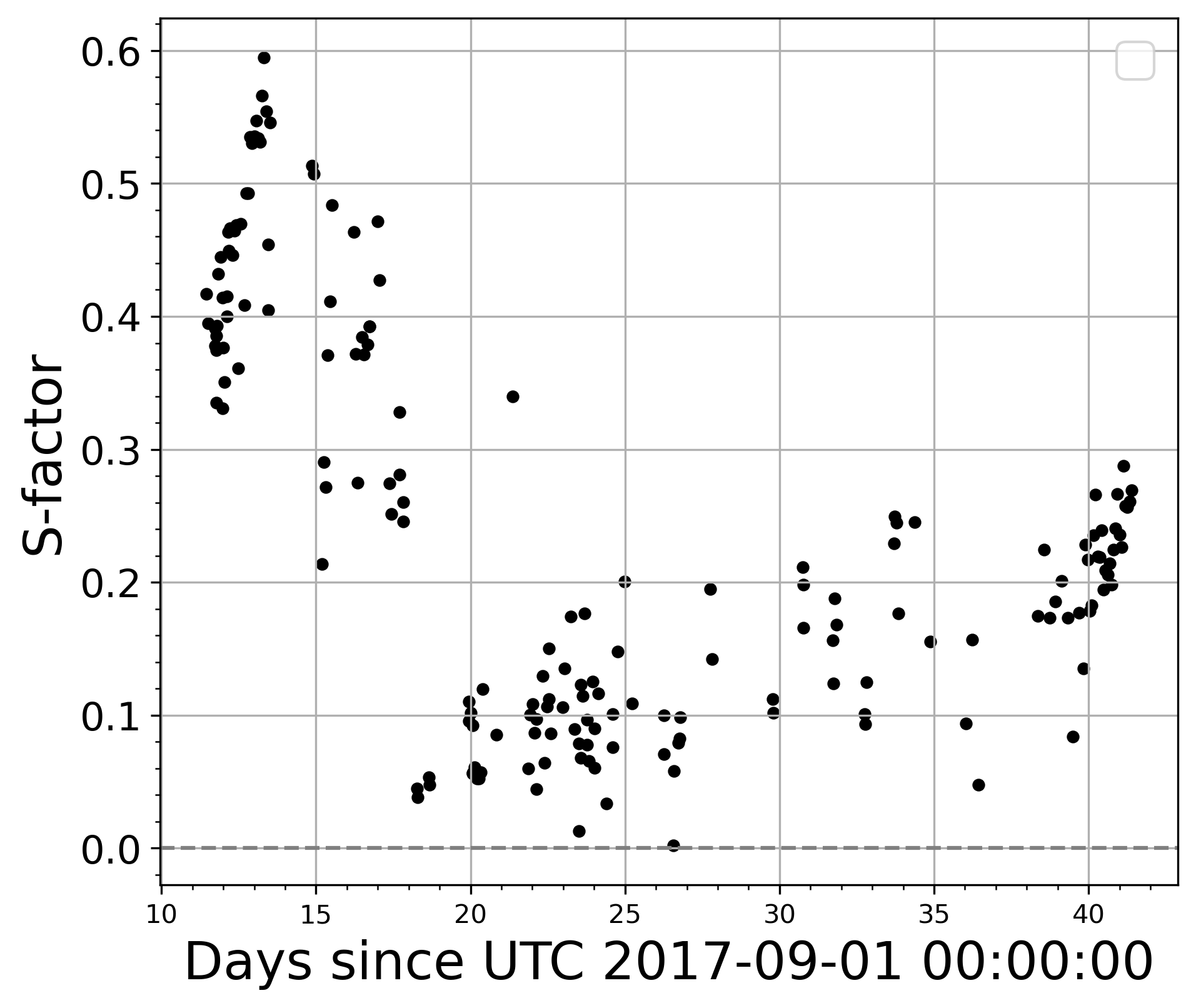}
	\end{minipage}
	\begin{minipage}{0.5\textwidth}
		\includegraphics[width=1\textwidth]{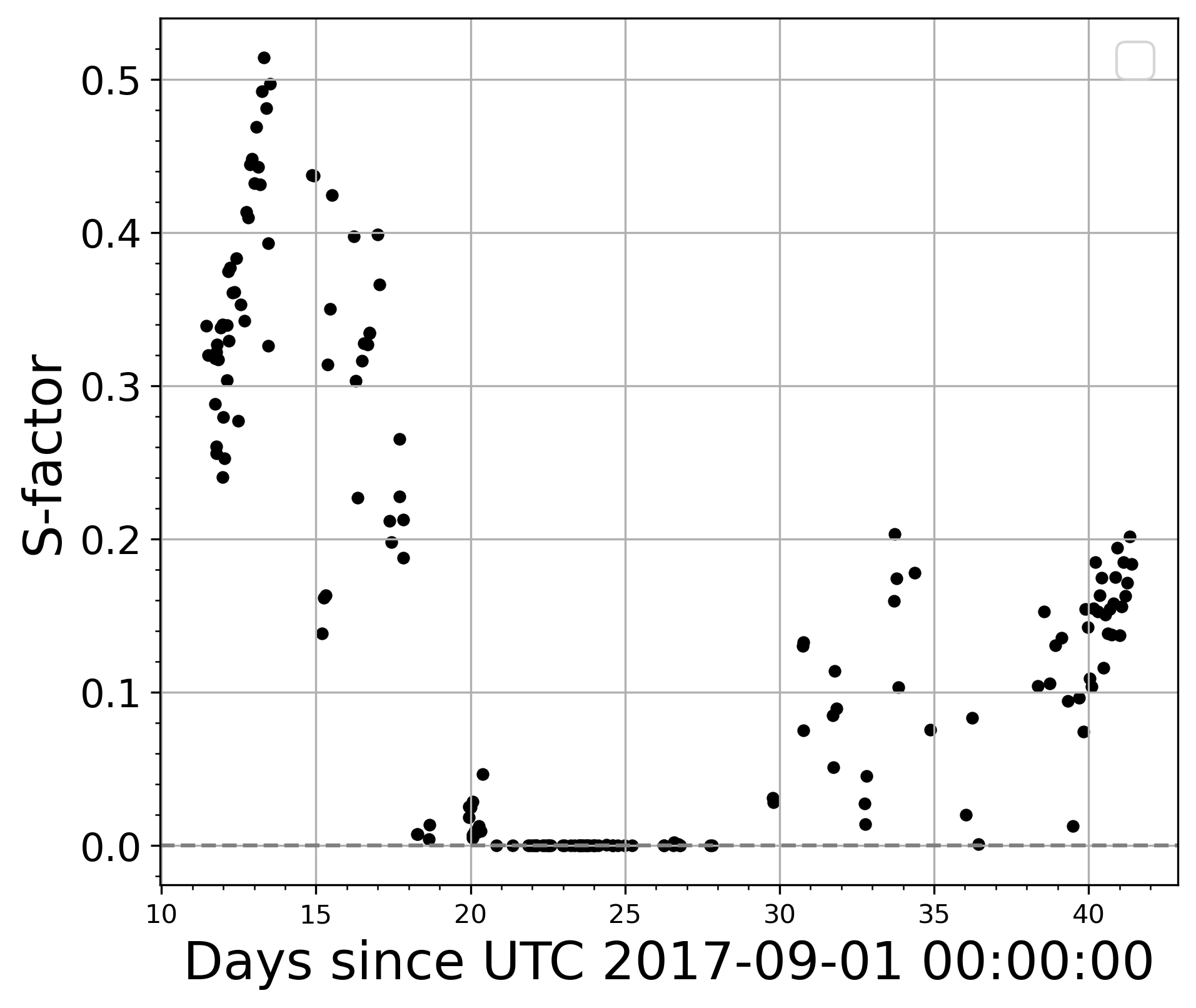}
	\end{minipage}
    \caption{S-factor evolution calculated with 95\% confidence level. {\bf Top panel:} without QPO component extraction in wavelet. {\bf Bottom panel:} after QPO component extraction in wavelet. GTI $> 100 s$.}
    \label{fig:S_compare}
\end{figure}

The S-factor, as defined in \cite{Chen2022MNRAS.517..182C}, is the ratio of the effective oscillation time and the total time in a selected GTI, where the effective oscillation time is defined as the following. Firstly, we define an "effective range" within the local wavelet spectrum, which represents the centroid frequency error range. Within this range, we exclude portions affected by the coi to ensure that we only consider areas with reliable signals. Then, for each time point, we examine its corresponding "effective range". If any power within the "effective range" at that time point exceeds the 95\% confidence level, it is labeled as an "effective oscillation time". The calculation of the error of S-factor is based on the NICER time resolution, which is quite small compared to S-factor, thus are ignored in our figures. 
In Figure~\ref{fig:S_compare}, we show the S-factor evolution with and without QPO component extracted in wavelet. The values drops overall by about 0.1 after the method applied. The shape of the evolution on the other hand, does not change too much except the type-B QPO time range as marked by \cite{Stevens2018ApJ...865L..15S} and \cite{Zhang2023MNRAS.520.5144Z}, which are almost always zero. The zero S-factor here means no significant power detected within the QPO frequency range, indicating that the signal of type-B QPO is very week or completely hidden beneath the BBN.

\begin{figure}
    \includegraphics[width=\columnwidth]{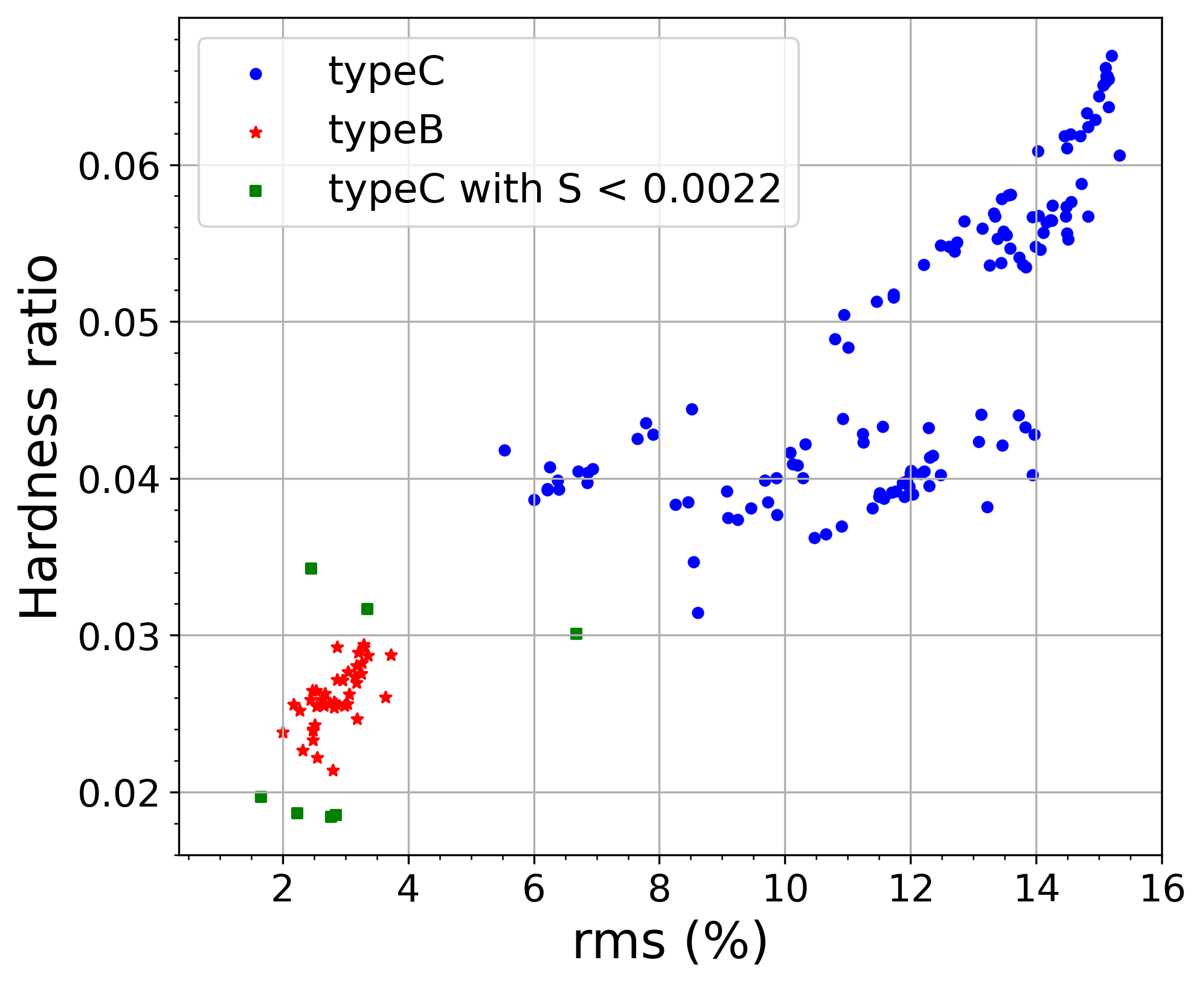}
    \caption{Hardness ratio (7-10 keV / 1-2 keV) and rms (3-10 keV in 1.5-15 Hz) relation. Blue points are type-C QPOs, green squares are type-C QPOs with S-factors less than 0.0022, and red stars are type-B QPOs.}
    \label{fig:hr-rms}
\end{figure}

\begin{figure}
    \includegraphics[width=\columnwidth]{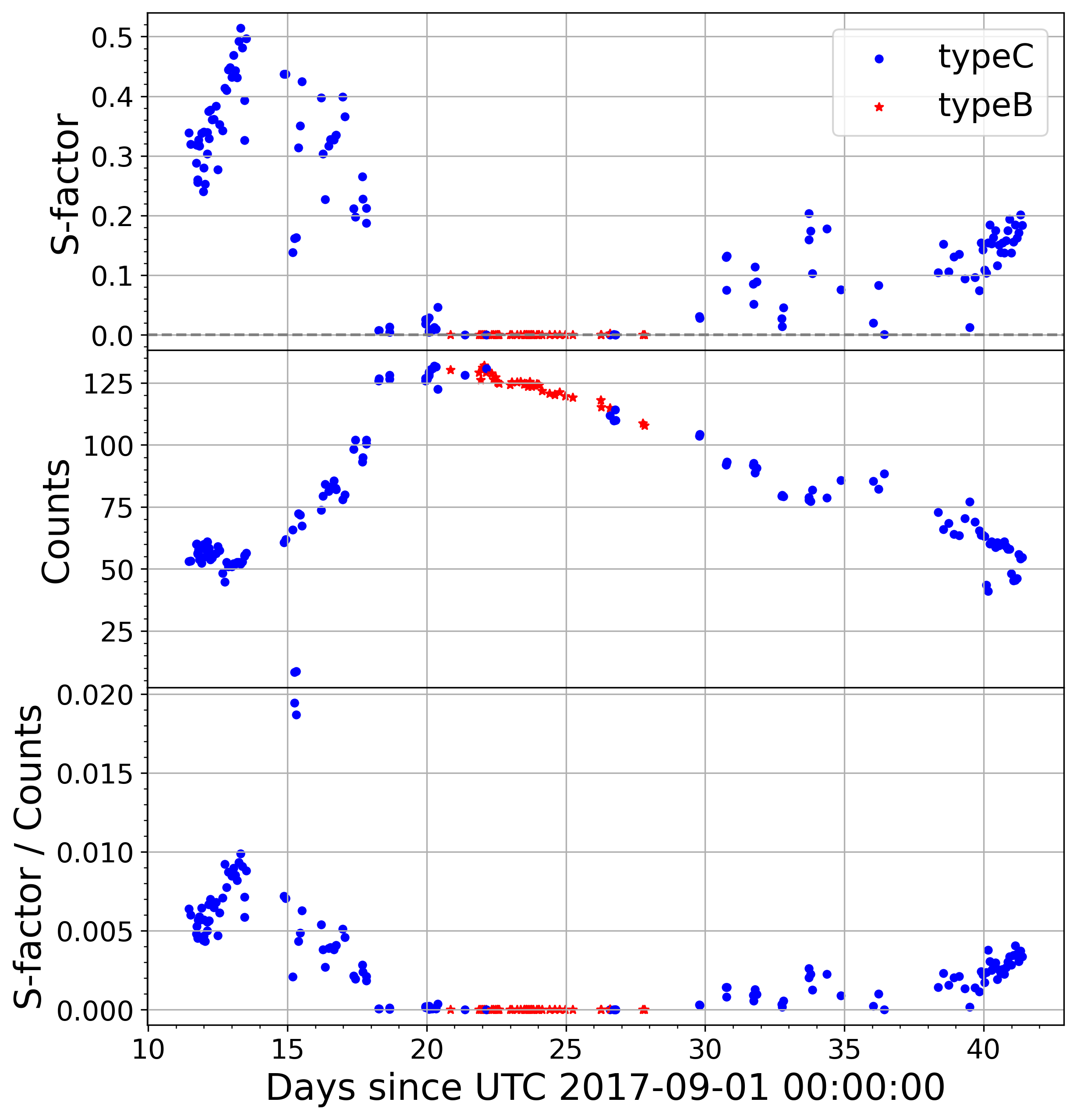}
    \caption{The evolution of S-factor calculated with 95\% confidence level (top panel), counts (middle panel) and the ratio of S-factor to counts (bottom panel). Blue points are type-C QPOs, while red stars are type-B QPOs.}
    \label{fig:S_3sub}
\end{figure}

\begin{figure}
    \includegraphics[width=\columnwidth]{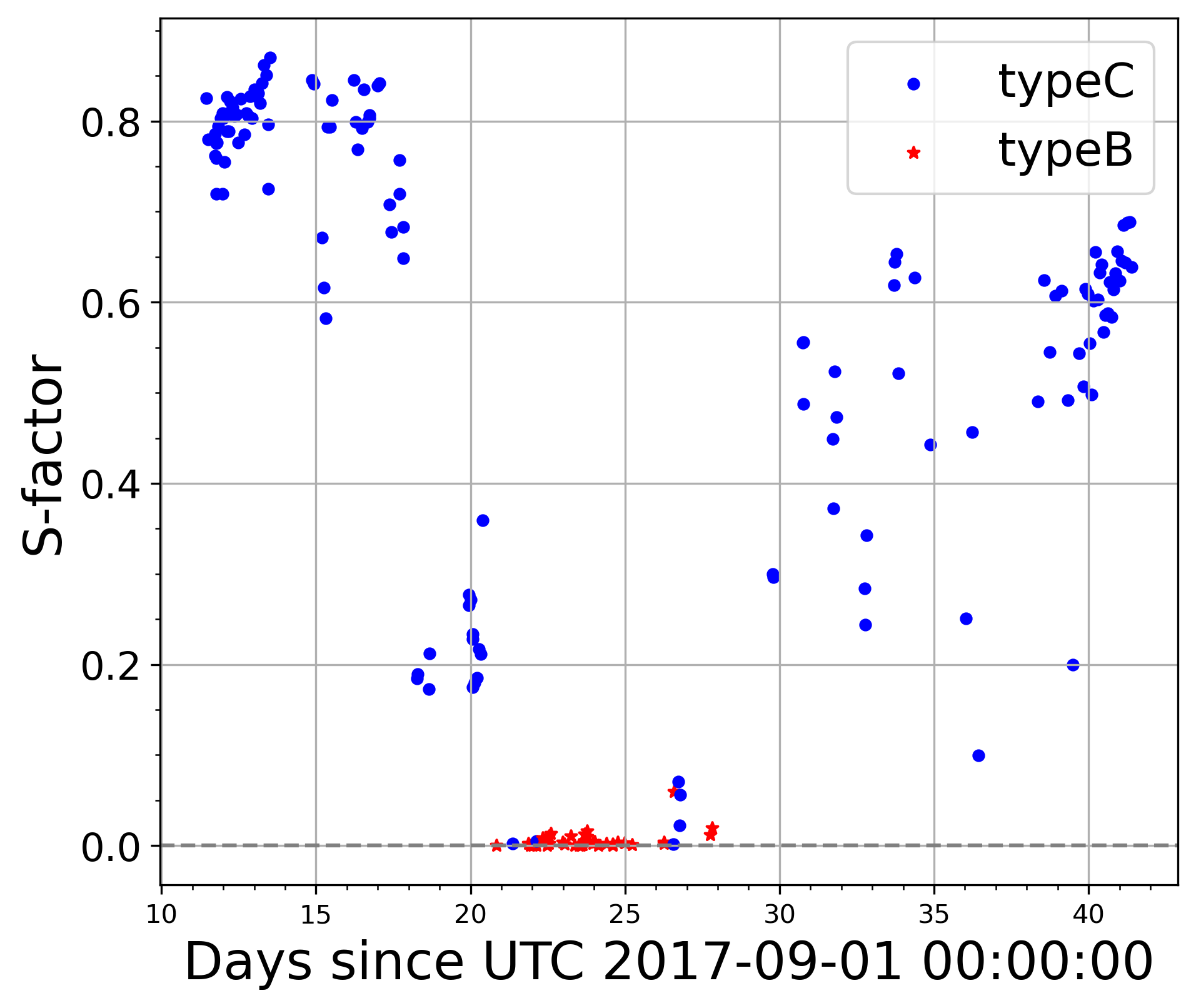}
    \caption{S-factor evolution calculated with 68\% confidence level. Blue points are type-C QPOs, while red stars are type-B QPOs.}
    \label{fig:S_sig68}
\end{figure}

To further verify the differences in the S-factor between type-B and type-C QPOs, we refer to a method provided by \cite{Stevens2018ApJ...865L..15S} to quantitatively separate them based on the range of hardness ratio and rms, where hardness ratio is defined as the ratio of 7-10 keV to 1-2 keV count rate, and rms in 3-10 keV is calculated in the frequency range of 1.5-15 Hz. They found that the type-B QPOs are in the range of 0.021-0.031 for hardness ratio, and less than 5 \% for rms in 3-10 keV. Based on these criterion, we re-generate the light curves to calculate the hardness ratio and rms in each GTI. The hardness ratio and rms results are shown in Figure~\ref{fig:hr-rms}. We then separate the QPO types and plot the evolution of S-factor on the top panel of Figure~\ref{fig:S_3sub}. All the type-B marked QPOs has the S-factor of zero except one GTI in 1050360119, which still shows a very small value of 0.0022. Among all the type-C marked QPOs, there are 7 points with S-factors smaller than this value, including one GTI in 1050360114, one GTI in 1050360115, four GTIs in 1050360119, and one more GTI in 1130360110, which are shown in Figure~\ref{fig:hr-rms} with green squares. Others are all greater or equal to 0.01. The former six points are all located very close to the type-B QPO region as reported in Figure 2 of \cite{Stevens2018ApJ...865L..15S}, showing rms $<$ 5\% and hardness ratio around 0.019 or 0.032. Considering \cite{Stevens2018ApJ...865L..15S} selected the hardness ratio region by sliding the window by 30\% increments, these six points may still be in the type-B QPO region. The last point in 1130360110 however, has rms of 0.067 and hardness ratio of 0.030. It is possible that rms selection performed by \cite{Stevens2018ApJ...865L..15S} was not quite rigorous, because the limit was chosen based on a gap between 4\% and 5.5\%. The 95\% confidence level result may ignore some QPO signals, thus a 68\% confidence level result is recalculated for comparison, and is shown in Figure~\ref{fig:S_sig68}. It is clear that all the S-factor of type-B QPOs are smaller than 0.1, including the seven points mentioned earlier. A distinct gap in S-factor can be noticed between them and other type-C QPOs, which normally have S-factor greater than 0.18. Considering the S-factor results of both Figure~\ref{fig:S_3sub} and Figure~\ref{fig:S_sig68}, the S-factor calculated through wavelet analysis might offer another reference parameter for the classification of QPO types that the S-factor of type-B QPO is notably smaller than that of type-C QPO. However, a more comprehensive study with different telescopes and sources will be absolutely required.


\begin{figure}
    \includegraphics[width=\columnwidth]{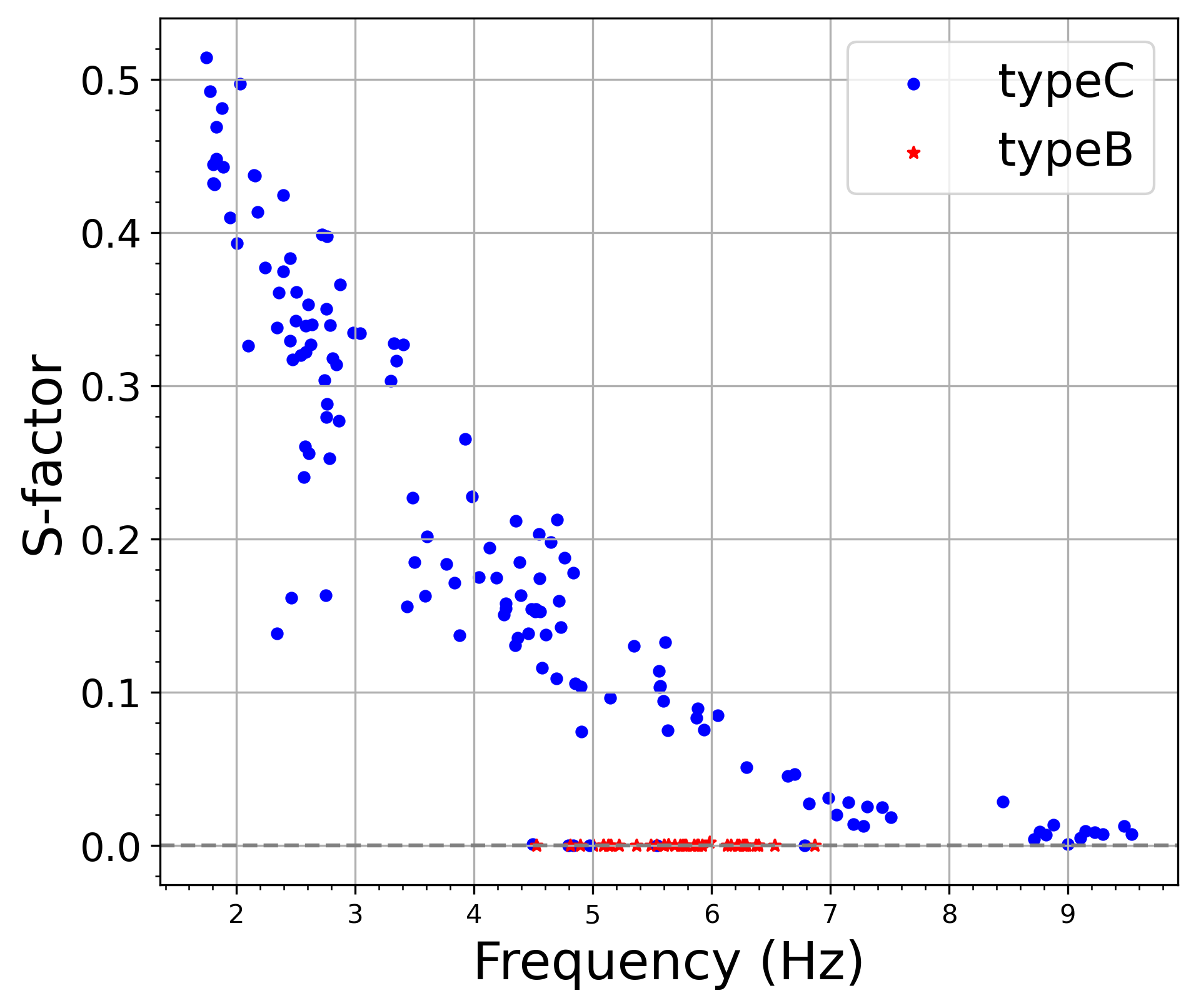}
    \caption{Frequency and S-factor (95\% confidence level) relation . Each point presents one GTI. Blue points are type-C QPOs, while red stars are type-B QPOs.}
    \label{fig:fs-relation}
\end{figure}

To further understand the properties of S-factor, we plot it with the QPO frequency in Figure~\ref{fig:fs-relation}, and find that it exhibits a very similar property with rms, which makes sense because they both represent the distribution properties of power, either in wavelet or in PDS. \cite{Motta2015MNRAS.447.2059M} investigated the relation of QPO rms and QPO centroid frequency to study the differences in high and low inclination sources. The anti-correlation of type-C QPO between rms and centroid frequency greater than 2 Hz can be noted, while the correlation of type-B QPO is not significant in their study. They are both similar to the results shown by the S-factor, and the different behavior of type-C and type-B QPOs may be discrete radio jet related \citep{Stevens2016MNRAS.460.2796S,deRuiter2019MNRAS.485.3834D}. Unfortunately, we do not have QPO centroid frequency below 1 Hz of MAXI J1535-571, otherwise we can show whether this source, as a potential high-inclination black hole binary, exhibits S-factor properties that are completely similar to QPO rms.

In addition, we notice that the relation of S-factor and count number might be energy dependent. As shown in Figure~\ref{fig:S_3sub}, when the count decrease in a small range at the very beginning, the S-factor increases significantly. Then the count increases and the S-factor decreases. When the outburst enters SIMS in which the type-B QPO dominated, the S-factor remains stable at zero. After that, the count number continue decreases and the S-factor raises again. The ratio of S-factor to counts does not remain constant for the type-C QPOs. Instead, it appears as a decrease in value in both sides when approaching the SIMS. These results show that, for type-C QPO, the S-factor and counts are inversely related, while the S-factor of type-B QPO does not change with count number. \cite{Chen2022MNRAS.517..182C} find the positive correlation of S-factors of type-C QPOs and count rates based on ME and HE bands of Insight-HXMT data, and the S-factor in LE band does not change with count rate. We speculate that the latter may be due to the fact that BBN accounts for a large proportion in the calculation of S-factor, since the QPO component is quite weak in LE after P011453500902. Nevertheless, due to the weak influence of BBN on the QPO component in the ME and HE bands, the QPO appears to perform different behaviors below and above 10 keV. A possible explanation is that the generation mechanism of QPO is different at lower energy band ($<$ 10 keV) which dominated by disc component, and at higher energy band ($> 10$ keV) dominated by corona.

Recent research shows that BH X-ray binaries may consist of two coronas, such as MAXI J1348$-$630 \citep{Garcia2021MNRAS.501.3173G} and GX 339$-$4 \citep{Peirano2023MNRAS.519.1336P}, by fitting spectra and type-B QPOs with a dual-component comptonization model. \cite{Peirano2023MNRAS.519.1336P} proposed that the inner hotter part of the disc provides the seed photons of the small corona, and the outer cooler part of the disc provides the seed photons of the large corona. \cite{Rawat2023MNRAS.520..113R} studied the same NICER observations of MAXI J1535-571 with a single-component time-dependent Comptonization model, but explained the difference between the temperature of the inner disk radius $kT_{in}$ and the temperature of the seed photon source $kT_s$ with a dual-corona geometry. If so, the different S-factor relation between lower band ($<10$ keV) and higher band ($10-100$ keV) may be related to the dual-corona geometry.


\section{Conclusions}

Wavelet analysis, with the QPO component extracted only, is utilized to study the QPO evolution in MAXI J1535-571 during its 2017 outburst. Based on the confident intervals, the light curve is separated into QPO time and non-QPO time, and significant differences are noticed in the PDS, hardness ratio and mean counts, indicating the QPO signals do not continuously show in the light curve of MAXI J1535-571 but appear and disappear on the order of seconds, which support the wavelet results observed by Insight-HXMT \citep{Chen2022MNRAS.513.4875C}. 

The wavelet method normally cannot be used to calculate the magnitude of power because of the mother wave utilized in the method, which will definitely smooth out the power. On the other hand, it can provide a more stable result with no "spiking noise" which is very common in the PDS. Hence, the parameters derived from wavelet such as the S-factor, which shows the proportion of time with significant QPO power, could be more stable. Therefore, using NICER data, we conduct a study on the S-factors of type-B and type-C QPOs in MAXI J1535-571, and find the existence of differences, which may contribute to the classification of QPOs. Additionally, we investigate the properties of the S-factor and find that it may be energy dependent and perhaps share similarities with rms. However, since wavelet analysis is rarely used in QPO research, the results regarding the S-factor require support from future research.


\section*{Acknowledgements}
We are grateful to the referee for the useful suggestions to improve the manuscript. 
This work is supported by the National Key Research and Development Program of China (Grants No. 2021YFA0718503), the NSFC (No. 12133007).

\appendix



\bibliographystyle{elsarticle-harv} 
\bibliography{example}






\end{document}